# PREPRINT DEPARTMENT OF PHYSICS AND ELECTRONICS, UNIVERSITY OF PUERTO RICO AT HUMACAO

# EARTH AND MARS CRATER-SIZE FREQUENCY DISTRIBUTION AND IMPACT RATES: THEORETICAL AND OBSERVATIONAL ANALYSIS

## W. BRUCKMAN, A. RUIZ, E. RAMOS

William Bruckman, Abraham Ruiz : *Department of Physics and Electronics*, And
Elio Ramos : *Department of Mathematics*, University of Puerto Rico At Humacao

## ABSTRACT

A framework for the theoretical and analytical understanding of the impact crater-size frequency distribution is developed and applied to observed data from Mars and Earth. The analytical model derived gives the crater population as a function of crater diameter, $D$, and age, $\tau$, taking into consideration the reduction in crater number as a function of time, caused by the elimination of craters due to effects such as erosion, obliteration by other impacts, and tectonic changes. When applied to Mars, using Barlow's impact crater catalog, we are able to determine an analytical curve describing the number of craters per bin size, $N(D)$, which perfectly reproduces and explains the presence of two well-defined slopes in the $\log[N]\ vs \log[D]$ plot. For craters with $D \leq \sim 57 km$, we find that $N\ \alpha\ 1/D^{1.8}$, and this distribution corresponds to 'saturation', where the rate of production is equal to the rate of destruction of craters. A steeper slope, with $N \alpha\ 1/D^{4.3}$ is found for the larger craters, $D \geq \sim 57 km$, from which it is interpreted that $N$ is pristine, or essentially unaffected by destructive mechanisms. Since in this limit of larger $D$ the rate of impacts, $\Phi(D)$, is proportional to $\frac{1}{D^{4.3}}$, and the cumulative rate, $\Phi_C(D) \equiv \int_D^\infty \Phi dD$, is proportional to $1/D^{3.3}$, we are able to estimate that the rate of meteorite impacts for energies, $E$, of a megaton or above ($D \geq \sim 1 km$) is about one every three years, a result that is relevant for future Mars explorations. The corresponding calculations for our planet give a probability of one per 15 years for an impact $E \geq \sim$megaton, while for a Tunguska–like event, where $E = 10$ megatons, an estimate of one per nearly a century is obtained. Our cumulative flux $\Phi_C$ is also expressed as a function of the meteorite diameter, $d$, and $E$, thus obtaining for Earth: $\Phi_C = \frac{5.5[1\mp0.5]}{10^6 years\ d^{2.57}} = \frac{[1\mp0.5]}{14.5 years\ E^{0.86}}$, where $d$ is in kilometers, and $E$ in megatons, results that are similar to those of Poveda et al (1999). The model allows the derivation of a more general expression, $\widetilde{N}$, that gives the number of craters observed today in the grid of diameters between $D$ and $D + \Delta D$ and with an age between $\tau$ and $\tau + \Delta \tau$. The application of $\widetilde{N}$ to describe the Earth's crater data shows a remarkable agreement between theory and observations.

## 1. Introduction and Summary

The present impact crater-size frequency distribution, $N(D)$, is the result, on one hand, of the rate of crater formation $\Phi$ attributed to the impacts of asteroids and comets, and, on the other hand, of the elimination of craters as they get older, by processes like erosion, obliteration by other impacts, tectonic changes, etc. Therefore, in order to understand the history of crater formation we will, in section 2, combine the above creation and elimination factors to show that we can express $N(D)$ in terms of $\Phi$ and the mean-life of a crater of diameter $D$, $\tau_{mean}$. Then, a simple model based on the above considerations is discussed in section 3, where we describe the crater-size distribution for planet Mars data, collected by Barlow (Barlow, 1988), and find that $\bar{\Phi} \alpha D^{-4.3}$ and $\tau_{mean} \alpha D^{2.5}$, where $\bar{\Phi}$ is the time average of $\Phi$.

In section 4 we use the above formalism to describe Earth's crater data as a function of crater age $\tau$, and conclude that on our planet it is also true that $\tau_{mean} \alpha D^{2.5}$. This interesting result is interpreted to mean that on the surface of both Mars and Earth, $\tau_{mean}$ is approximately proportional to the volume of the crater $\equiv D^2 h$, where $h$ is defined as the average height of the crater and accordingly satisfies $h \alpha D^{1/2}$. Investigations of the geometric properties of a Martian impact crater (Garvin 2002) indeed confirm that $h(D) \cong D^{0.5}$.

Next, in section 5, we take on the calculation of $\Phi(D)$ and $\Phi_C(D)$ for our planet, where $\Phi_C(D)$, 'the cumulative flux', is the rate of impact of craters with diameters larger than $D$. $\Phi_C$ is also expressed as a function of the impactor diameter $d$ and its kinetic energy $E$. For instance, our results predict one megaton or larger energy bolide with probable period of about one every $15/[1 \mp 0.5]$ years, while nearly one Tunguska-like event ($E \geq 10$ megatons) is expected every century: $105/[1 \mp 0.5]$ years. Another interesting result is that, according to our model, a meteorite impact of catastrophic potential, $D \geq \sim 5km$, $d \geq \sim 0.2km$, is likely to have occurred within historical times: $\tau \leq \sim 5,000$ years. Generally speaking, we find that for our planet we have:

$$\Phi_C = 2.8\left[\frac{1 \mp 0.50}{my}\right](20/D)^{3.3} = \frac{5.5[1 \mp 0.5]}{my\ d^{2.57}} = \frac{[1 \mp 0.5]}{14.5y\ E^{0.86}},$$

where $d$ and $D$ are in kilometers, $E$ in megatons, and $my \equiv 10^6$ years $\equiv 10^6 y$. The above result is in almost perfect agreement with the results of Poveda et al (1999). We compare, in section 6, the fluxes of Earth and Mars, and conclude that astronauts on Mars will probably be subjected to a meteorite with at least a megaton of energy in a mission lasting more than about three years. This intriguing result raises concerns for manned missions to the red planet in the near future.

In the last section we consider an application of a general expression $\tilde{N}(D, D + \Delta D, \tau, \tau + \Delta\tau)$ to Earth, that gives the number of craters observed today in the grid of diameters between $D$ and $D + \Delta D$ and with an age between $\tau$ and $\tau + \Delta\tau$. First, in Table III and Figure (5), $\tilde{N}(\tau)$ is calculated for craters older than $\tau$, $1my \leq \tau \leq 2,500my$, and for all diameters $\geq 20km$, and this is compared with the observations. The crater data was obtained from http://www.passc.net/EarthImpactDatabase/, where we used $D \geq 20km$ and the following selected area to minimize the undercounting of craters: Australia, Europe, Canada up to the Arctic Circle latitude and the United States. We found that the theoretical $\tilde{N}(\tau)$ remarkably reproduces the observed data. Furthermore, in Table IV and Figure (6) we again compare the theory and observations, this time for craters of all ages but with $\tilde{N}(diameters \geq D)$ vs. $D$, ($1km \leq D \leq 300km$). Here we see, as expected, a discrepancy with observational data for $D \leq \sim 20km$, due to undercounting, but a very good agreement for $D \geq 20km$.

## 2. Theoretical Model for the Observed Data

In what follows, we will present a theoretical and analytical curve that will reproduce the essential features of the Martian crater-size frequency distribution empirical curve (Figure (1)), based on Barlow's (1988) database of about 42,000 impact craters. The models will be derived using reasonably simple assumptions, which will allow us to relate the present crater population to the crater population at each particular epoch.

To this end, let $\Delta N(D,t)\Delta D$ represent the number of craters of diameter $D \mp \Delta D/2$ formed during the epoch $t \mp \Delta t/2$, where we are assuming that $\Delta t$ and $\Delta D$ are sufficiently large to justify treating $\Delta N$ as a statistically continuous function, but sufficiently small to be able to treat them as differential in the following discussion. This initial population will change as time goes on due to climatic and geological erosion and the obliteration of old craters by the formation of new ones. Therefore, we expect that the change in $\Delta N$ during a time interval $dt$ will be proportional to itself and $dt$:

$$d(\Delta N) = -C\Delta N dt \qquad (1)$$

where $C$ is a factor that takes into account the depletion of the craters, and should be a function of the diameter, since the smaller a crater is, the more likely it is to disappear. It is easy to integrate Eq. (1) in time to obtain

$$\Delta N(D,t) = \Delta N(D,t_n) Exp[-\bar{C}\tau_n], \qquad (2)$$

$$\bar{C}\tau_n = \int_{t_n}^{t} C \, dt`, \qquad (3)$$

$$\tau_n \equiv t - t_n, \qquad (4)$$

which is the familiar equation for an exponential decay of $\Delta N$ with mean-life: $\tau_{mean} = 1/\bar{C}$.

Eq. (2) gives the number of craters, per bin size, as a function of diameter, that are observed at time $t$ but were produced in the time interval $t_n \mp \Delta\tau_n/2$. Therefore, the total contribution to the present ($t = 0$) population of craters is:

$$N(D,0) \equiv N(D) = \sum_n \Delta N(D,t_n) Exp[-\bar{C}\tau_n] = \sum_n [\Delta N(D,t_n)/\Delta\tau_n] Exp[-\bar{C}\tau_n]\Delta\tau_n. \qquad (5)$$

Alternatively, in the continuous limit $\Delta\tau_n \to d\tau \to 0$, Eq. (5) becomes

$$N(D) = \int_0^{\tau_f} \{\Phi(D,\tau) Exp[-\bar{C}\tau]\} d\tau, \qquad (6)$$

where

$$\Phi(D,\tau) \equiv \lim_{\Delta\tau_n \to 0}(\Delta N(D,\tau_n)/\Delta\tau_n). \qquad (7)$$

We see that $\Phi(D,\tau)$ corresponds to the rate of formation of craters per bin size, of diameter $D$ in the epoch $\tau$, and $\tau_f$ is the total time of crater formation.

Another derivation of Eq. (6) is as follows. The number of craters formed over time $dt$ is $\Phi dt$, while the craters eliminated in this time are $CNdt$. Therefore, the net change in $N$ is

$$dN = \Phi dt - CNdt, \tag{8}$$

which is an equation that can be integrated by an elementary method, illustrated below. Multiplying the above equation by $Exp[-\int_0^\tau C\,d\tau`] \equiv Exp[-\bar{C}\tau]$, with $\tau = -t,$ we arrive at

$$Exp[-\bar{C}\tau](dN - CNd\tau) = -Exp[-\bar{C}\tau](\Phi d\tau), \tag{9}$$

which, after using

$$d(\bar{C}\tau) \equiv d\int_0^\tau C\,d\tau` = Cd\tau, \tag{10}$$

becomes

$$d(N Exp[-\bar{C}\tau]) = -Exp[-\bar{C}\tau](\Phi d\tau). \tag{11}$$

Integrating Eq. (11) from $0$ to $\tau_f$ with $N(\tau_f) = 0$ gives Eq. (6). Note that if, instead of the time interval $\tau = 0$ to $\tau = \tau_f$ in Eq. (6), we have the arbitrary limits $\tau$ to $\tau + \Delta\tau$, we obtain the more general expression:

$$N(D,\tau,\tau+\Delta\tau) = \int_\tau^{\tau+\Delta\tau}\{\Phi(D,\tau`)\,Exp[-\bar{C}\tau`]\}d\tau`. \tag{12}$$

The $N$ defined in Eq. (6) represents the number of craters of diameter $D$ per bin size observed today ($t = 0$) that are younger than age $\tau_f$, while in Eq. (12) $N$ represents the number of craters observed at $t = 0$ formed, per bin, in the time interval $\tau$ to $\tau + \Delta\tau$. Accordingly, the number of present craters in an arbitrary diameter interval $D$ to $D + \Delta D$ with age between $\tau$ and $\tau + \Delta\tau$ is represented by the integral:

$$\tilde{N}(D, D+\Delta D, \tau, \tau+\Delta\tau) \equiv \int_D^{D+\Delta D} N(D,\tau,\tau+\Delta\tau)dD. \tag{13}$$

For instance, if $\Delta D \to \infty$, the integral is called the total cumulative number of craters. Further discussions and applications of Eq. (13) to the Earth's crater record will be continued in section 7. For planet Mars, however, we will be applying Eq. (6) in the next section.

## 3. Applications to Mars

In Appendix A we show that Eq. (6) can be rewritten in the form:

$$N = (\overline{\Phi/C})\{1 - Exp[-\bar{C}\tau_f]\}, \qquad (14)$$

where

$$\overline{(\Phi/C)} \equiv \frac{\int_0^\tau (\Phi(\tau`)/C)\{Exp[-\int_0^{\tau`} C(\tau``)d\tau``]\}Cd\tau`}{\int_0^\tau \{Exp[-\int_0^{\tau`} C(\tau``)d\tau``]\}Cd\tau`}. \qquad (15)$$

Notice that $\overline{(\Phi/C)}$ is the weighted average of ($\Phi/C$), with weight:

$$\int_0^\tau \{Exp[-\int_0^{\tau`} C(\tau``)d\tau``]\}Cd\tau`. \qquad (16)$$

We now consider Barlow's (1988) data for Mars' crater-diameter distribution, which is plotted in Figure (1). Then the simplest model that describes these data, for $D \geq 8km$, is given by the expressions (Figure (2)):

$$\overline{(\Phi/C)} = \frac{1.43 \times 10^5}{D^{1.8}}, \qquad (17)$$

$$\bar{C} = \left(\frac{1}{\tau_f}\right)\frac{2.48 \times 10^4}{D^{2.5}}, \qquad (18)$$

$$\bar{\Phi} = \left(\frac{1}{\tau_f}\right)\frac{3.55 \times 10^9}{D^{4.3}}. \qquad (19)$$

$$\overline{(\Phi/C)} = \frac{\bar{\Phi}}{\bar{C}} \qquad (20)$$

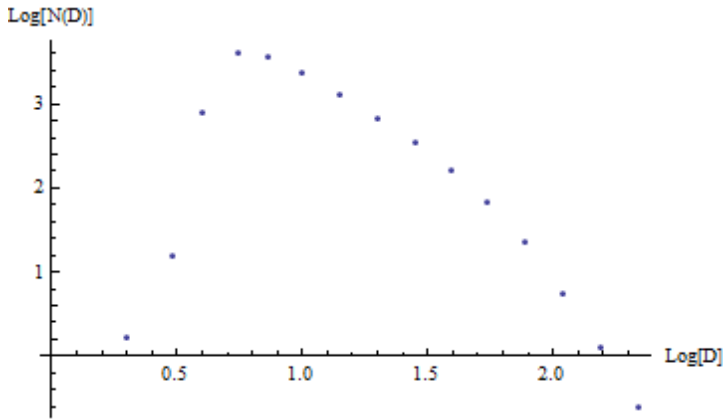

FIGURE (1): Log-Log plot of number of craters per bin, $N(D)$ vs $D$ based on Barlow's Mars catalog. The number $N(D)$ is calculated by counting the number of craters in a bin $\Delta D = D_R - D_L$, and then dividing this number by the bin size. The point is placed at the mathematical average of $D$ in the bin: $(D_R + D_L)/2$. The bin size is $\Delta D = (\sqrt{2} - 1)D_L$, so that $\frac{D_R}{D_L} = \sqrt{2}$.

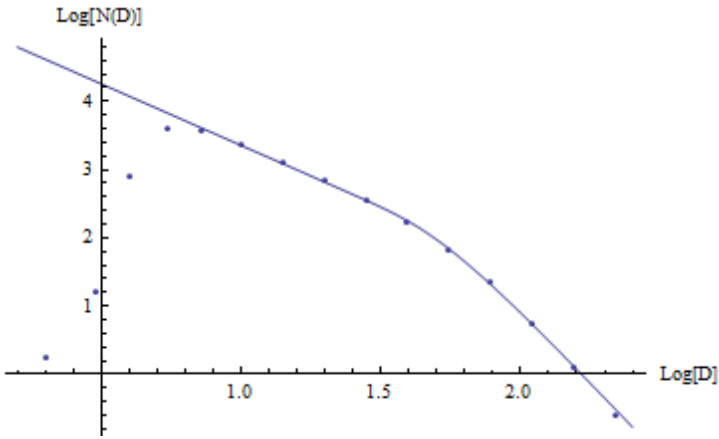

FIGURE (2): Comparing the model in Eqs. (14) to (20) with the Mars data in Figure (1).

We see that the theoretical curve shown in Figure (2) differs significantly from the observed data for $D$ less than about $8km$. However, according to Barlow, her empirical data undercounts the actual crater population for $D$ less than $8km$ and therefore, we will restrict our analysis to $D \geq 8km$.

Eqs. (2) and (18) imply that the fraction of craters of diameter $D$ formed at each epoch $\tau$ that still survive at the present time $t = 0$ is given by:

$$\frac{\Delta N(D,0)}{\Delta N(D,\tau)} = Exp[-\bar{C}\tau] \approx Exp[-(57/D)^{2.5}\frac{\tau}{\tau_f}] \tag{21}$$

and thus, the mean life for craters of diameter $D$, $\tau_{mean} \equiv \bar{C}^{-1}$, is:

$$\tau_{mean} \equiv \bar{C}^{-1} \approx (D/57)^{2.5}\, \tau_f. \tag{22}$$

Hence, craters with $D \approx 57km$ have $\tau_{mean} \approx \tau_f$, whereas

$$\tau_{mean} \gg \tau_f,\ D \gg 57km, \tag{23}$$

$$\tau_{mean} \ll \tau_f,\ D \ll 57km. \tag{24}$$

In the limit $D \gg 57km, \bar{C}\tau_f \ll 1$, we obtain, from Eqs. (14),(20) and (19), that:

$$N = \bar{\Phi}\tau_f = \frac{3.55 \times 10^9}{D^{4.3}};\ \bar{C}\tau_f \ll 1,\ D \gg 57km, \tag{25}$$

which corresponds to a straight line of slope -4.3 in the $\log(N)$ vs. $\log(D)$ plot, which we see in the right-hand part of Figure (2), and is the form of Eq. (14) when we can ignore the destruction of craters ($\bar{C}\tau_f \ll 1$). In other words, for these larger craters, their number is simply given by:

$$N = \bar{\Phi}\tau_f \equiv \int_0^{\tau_f} \Phi d\tau, \tag{26}$$

an expected relationship when craters are conserved and therefore, when the actual crater number is proportional to the age of the underlying surface $\tau_f$. On the other hand, for smaller craters where $Exp[-\bar{C}\tau_f] \ll 1$ we will have, from Eqs. (14) and (17),

$$N = (\overline{\Phi/C}) = \frac{\overline{\Phi}}{\overline{C}} \equiv \overline{\Phi}\tau_{mean} = \frac{1.43 \times 10^5}{D^{1.8}}, \quad (27)$$

and hence in this limit, $N$ is proportional to the survival mean life, $\tau_{mean}$, of craters of size $D$. This feature was called the 'crater retention age' by Hartmann (2002) and on Mars is shown in craters with $D$ less than about $57 km$, corresponding to the straight line segment on the left-hand side of Figure (2) with slope -1.8. When this condition arises, $N(D)$ is independent of $\tau$, since crater production, $\Phi dt$, is balanced by crater destruction, $NCd\tau$, and thus, from Eq. (8) we have:

$$dN = \Phi dt - CN dt = 0, \quad (28)$$

or, equivalently,

$$N = \frac{\Phi}{C}. \quad (29)$$

We see that the time average of Eq. (29) gives rise to Eq. (27). Therefore, the above model tells us that the empirical curve is essentially constructed by the following two straight lines in the $logN(D)$ vs $logD$ plot:

$$N(D) = \frac{\overline{\Phi}}{\overline{C}} = \frac{1.43 \times 10^5}{D^{1.8}}; \quad \overline{C}\tau_f \geq 1, D \leq \sim 57 km, \quad (30)$$

$$N(D) = \overline{\Phi}\tau_f = \frac{3.55 \times 10^9}{D^{4.3}}; \quad \overline{C}\tau_f \leq 1, D \geq \sim 57 km. \quad (31)$$

The exponent 4.3 is pristine, while the exponent 1.8 is the result of a steady state equilibrium between elimination and creation of craters. The large exponent, 4.3, has interesting implications for the corresponding impactors size-frequency distribution, and we will elaborate on this topic in section 5.

## 4. Determination of the Mean Life, $\tau_{mean}$, of Impact Craters on Earth

In what follows, we will determine the mean life for craters on our planet. Let us first consider the expression defining the average diameter, with ages in a given bin time interval $\Delta\tau = \tau_R - \tau_L$, which is given by:

$$\overline{D} = \frac{\int_0^\infty D N(D,\tau_L,\tau_R) dD}{\int_0^\infty N(D,\tau_L,\tau_R) dD}, \quad (32)$$

where, according to Eq. (12),

$$N(D, \tau_L, \tau_R) \equiv \int_{\tau_L}^{\tau_R} \Phi(D,\tau)\{Exp[-\overline{C}\tau]\}d\tau. \quad (33)$$

Investigations of the time dependence of the cratering rate of meteorites have concluded (see, for example, Hartmann (1966); Neukum et al (1983); Neukum (2001); Ryder (1990)) that the Earth went through a heavy bombardment era and that the impact rate then decayed exponentially until about 3,000 to 3,500 million years ago and since that time has remained nearly constant until the present. Therefore, for the Earth's data which is younger than 3 to 3.5gy, we can reasonably assume that $\Phi$ is independent of $\tau$. Furthermore, following the analysis of Mars, we also assume that $\Phi$ and $\bar{C}$ are given by the simple polynomial forms

$$\Phi = \frac{A}{D^m} \; ; A, \; m \text{ are constants}, \tag{34}$$

$$\bar{C} = \frac{B}{D^p} \; ; B \text{ and } p \text{ are constants}. \tag{35}$$

Then we can show (see Appendix C) that with $g \equiv \tau_L/\tau_R$ and $m` \equiv m - p$, Eq. (32) becomes

$$\bar{D} = (B\tau_L)^{\frac{1}{p}} \alpha 1, \tag{36}$$

where:

$$\alpha 1 = \frac{\Gamma((m`-2)/p)}{\Gamma((m`-1)/p)} \frac{[1-g^{(m`-2)/p}]}{[1-g^{(m`-1)/p}]}, \tag{37}$$

and:

$$\Gamma(n) \equiv \int_0^\infty x^{n-1} e^{-x} dx \tag{38}$$

is the complete Gamma function. Eq. (36) can be rewritten as

$$\log[\bar{D}] = \left(\frac{1}{p}\right) \log[\tau_L] + \log[\alpha 1 B^{\frac{1}{p}}], \tag{39}$$

which is the equation for a straight line in the $\log[\bar{D}] \, vs \, \log[\tau_L]$ graph, with a slope of $1/p$ and an intercept $\log[\alpha 1 B^{\frac{1}{p}}]$. In Figures (3) and (4) we plot $\log[\bar{D}] \, vs \, \log[\tau_L]$, for $g = \frac{1}{2}$ and $g = 1/10$ respectively, using the Earth's crater size data against $\tau$. From Figure (3) we obtain $\frac{1}{p} \approx 0.39, B \approx 11/my$ while from Figure (4), $\frac{1}{p} \approx 0.43, B \approx 12/my$. The values $0.39$ and $0.43$ are very close to the value found for Mars: $\frac{1}{2.5} = 0.40$, and this result is interpreted as follows. If we assume that, as expected, $\tau_{mean}$ is a function of the volume of the crater, $V$, so that $\tau_{mean}$ increases with increasing $V$, then it is reasonable to use a Taylor series, and expand $\tau_{mean}$ in terms of $V$ to obtain:

$$\tau_{mean} \equiv \frac{1}{\bar{C}} = \sum k_i V^i = k_1 V + k_2 V^2 + \cdots. \tag{40}$$

Furthermore, in the linear limit where we keep only the first term in Eq. (40), we have:

$$\tau_{\text{mean}} \equiv \frac{1}{\bar{C}} \cong k_1 V \, \alpha \, D^2 h, \tag{41}$$

where $V \equiv D^2 h$, with $h$ defined as the average height of a crater of size $D$. The above equation could also be derived heuristically by arguing that the average change in crater volume is approximately proportional to time, $dV \, \alpha \, dt$, and thus, $V \cong \alpha \, \tau_{mean}$. Then with $p = 2.5$, Eqs. (41) and (35) imply that

$$h \cong \text{constant} D^{1/2}, \tag{42}$$

which is a prediction that was investigated, and we have found that Eq. (42) is indeed consistent with results from studies of impact crater geometric properties on the surface of Mars by J.B. Garvin (Garvin,2002).

Therefore, it appears that the Earth's crater distribution, as for Mars, satisfies the relationship:

$$\tau_{mean} \equiv \frac{1}{\bar{C}} \cong constant\, D^{2.5}, \tag{43}$$

and this simple behavior follows from the relation:

$$\tau_{mean}\; \alpha\; V = D^2 h, \text{ with } h\; \alpha\; D^{1/2}. \tag{44}$$

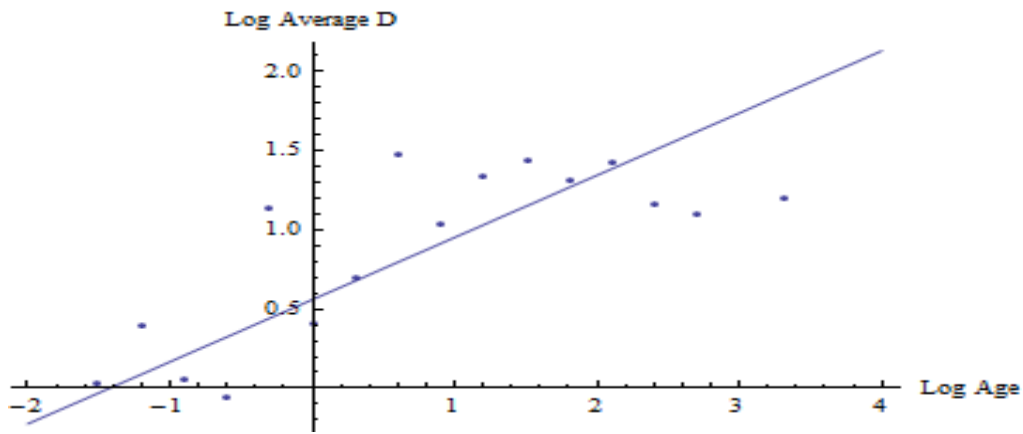

FIGURE (3): $[\bar{D}]$ vs $Log[\tau]$, with $g = 1/2$.

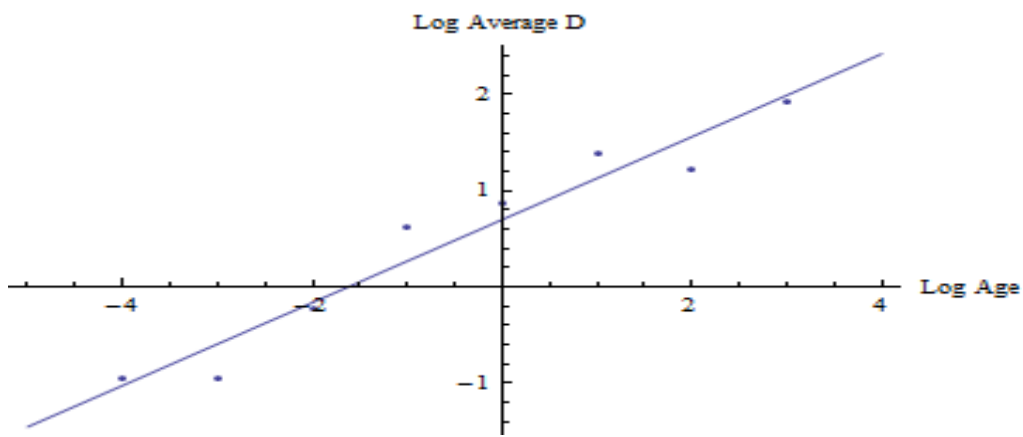

FIGURE (4): $[\bar{D}]$ vs $Log[\tau]$, with $g = 1/10$.

## 5. The Cumulative Flux on Earth

Let us now study the implications for our planet of a flux of the form:

$$\Phi = \frac{A}{D^{4.3}}, \tag{45}$$

corresponding to the cumulative flux:

$$\Phi_C(D) = \int_D^\infty \Phi dD = \frac{1}{3.3}\frac{A}{D^{3.3}}. \tag{46}$$

The value of $A$ can be estimated for Earth from the result of Grieve and Shoemaker (1994) for $D = 20km$:

$$\Phi_C(20km) = \frac{(5.5 \mp 2.7)10^{-9}}{(my)km^2} 4\pi R^2 \approx 2.8[\frac{1 \mp 0.50}{my}], \tag{47}$$

where $R$ is the Earth's radius, and $my$ is million years. Comparing Eq. (47) with Eq. (46) we obtain:

$$A = 9.24[1 \mp 0.50]\frac{(20)^{3.3}}{my}, \tag{48}$$

and thus we have:

$$\Phi_C(D) = 2.8[\frac{1 \mp 0.50}{my}](20/D)^{3.3}. \tag{49}$$

Table I illustrates the outcomes of the above formula for selected values of $D$. Note that $\Phi_C$ is the probable frequency of impacts, and $1/\Phi_C$ is the probable period between impacts, for diameters larger than or equal to $D$. It is interesting that $D = 5km$ has a statistical periodicity of about one in 3,680 years, suggesting that these are potentially historical events.

**Table I**

| D(KM) | $\Phi_{Acc}(D)/(1 \mp 0.50)$ | $[\Phi_{Acc}(D)/(1 \mp 0.50)]^{-1}$ |
|---|---|---|
| 200 | 1/(712my) | 712my |
| 150 | 1/(275my) | 275my |
| 100 | 1/(72my) | 72my |
| 50 | 1/(7my) | 7my |
| 10 | 28/(my) | 35,700y |
| 5 | 272/(my) | 3,680y |

The formation of craters with potential diameters of less than approximately $5km$ is strongly affected by the Earth's atmosphere, since these bodies can be fragmented or even disintegrated. Therefore, for $D < 5km$ we prefer to express the flux in terms of the kinetic energy, $E$, and the diameter, $d$, of the impactor. To convert $D$ to $d$ we will use the Schmidt and Holsapple scaling equation (Gault (1974) ,Holsapple (1987),Schmidt (1987),Melosh (1989)),which can be well approximated by the following expression (see, for example, Hughes (1998) and Ward (2002)):

$$D=10^{1.21a_1} d^{0.78a_2}, \tag{50}$$

where the values $a_1$ and $a_2$ are very close to 1, and so from now on we will put them as equal to 1. Substituting Eq. (50) in Eq. (49), we obtain:

$$\Phi_C = \frac{5.5[1\mp 0.5]}{my\, d^{2.57}} = \frac{(2.8)10^2}{yd_m^{2.57}}[1 \mp 0.5], \tag{51}$$

where $d$ is in kilometers and $d_m$ in meters. We can also express $\Phi_C$ in terms of the kinetic energy of the bolide, $E = \frac{1}{2}mv^2$, to get:

$$\Phi_C(E) = \frac{[1\mp 0.5]}{14.5y\, E^{0.86}} = \frac{26}{yE_{kt}^{0.86}}[1 \mp 0.5], \tag{52}$$

where $E$ is in megatons and $E_{kt}$ is in kilotons. To convert $d$ to $E$, we follow Poveda et al (1999) and thus use:

$$E = \left(\frac{1}{2}\right)Mv^2 = (4\pi\rho/6)\left(\frac{d}{2}\right)^3 v^2. \tag{53}$$

with $\rho = 2{,}400 kg/m^3$ and $v = 20 km/s$. Table II below gives values of $1/\Phi_C(E)$ for selected $E$ and corresponding approximate values of $d_m$ and $D$.

**Table II**

| $E$(megatons) | Approximate $d_m$ | Approximate $D(km)$ | $[1 \mp 0.5]/\Phi_C$ |
|---|---|---|---|
| 250 | 160 | 4 | 1673 years |
| 100 | 120 | 3 | 761 " |
| 50 | 90 | 2.5 | 419 " |
| 20 | 70 | 2 | 191 " |
| 10 | 60 | 1.7 | 105 " |
| 5 | 40 | 1.4 | 58 " |
| 2 | 30 | 1.1 | 26 " |
| 1 | 26 | 0.93 | 14.5 " |
| 0.1 | 12 | 0.51 | 2 " |
| 0.02=20kt | 7 | 0.34 | ½ year |
| 0.01=10kt | 5.7 | 0.28 | 3.3 months |
| 0.005=5kt | 4.5 | 0.24 | 1.9 " |
| 0.001=1kt | 2.6 | 0.16 | 2 weeks |

It is interesting to compare our results with those of Poveda et al (1999) which give:

$$\Phi_C(d)(Poveda) = \frac{L}{d_m^{2.5}}, \tag{54}$$

$$\Phi_C(E)(Poveda) = \frac{R}{E_{kt}^{0.83}}, \tag{55}$$

with exponents strikingly similar to our model. The values for $L$ and $R$ depend on three possible scenarios for the mean albedo composition of asteroids, and Poveda et al (1999) considered the following distribution for the albedo of asteroids:

Case I:   $L = (2.25)10^2/y, R = 18.74/y$

when 50% of asteroids have albedo 0.155(S-type) and 50% have albedo 0.034(C-type);

Case II:   $L = (1.6)10^2/y, R = 14.36/y$ ; for 70% S-type and 30% C-type

Case III:  $L = (3.75)10^2/y, R = 29.51/y$; for 30% S-type and 70% C-type

Our results from Eqs. (51) and (52) are $L = (2.8)10^2[1 \mp 0.5]/y,$ and $R = 26[1 \mp 0.5]/y$ respectively, which are remarkable close to the average values of the three distributions above, and within the uncertainties of our model.

It is worth pointing out that observations (see Silber et al (2009)) for megaton-size impacts give a frequency of about one every 15 years, which coincides with the corresponding value in our calculations. On the other hand, their estimated impact rate for energies of about 11-12 kt is approximately one per year, which is near 1/3 of our estimate, but Lewis (1996) concluded instead that the defense support program satellite data imply a rate of about 12 per year for these energies, which is about three times our estimate. However these discrepancies are still within the error bars. For the Tunguska type impact energy (about 10 megatons), we predict an accumulative rate of one in about 100 years, which is much higher than previous estimates. However, we are in very good agreement with Poveda et al (1999), and are not inconsistent with the estimates of Archer et al (2005), who put the period as one event in less than about 300 years.

## 6. Comparing Mars and Earth`s Impact Rates

We see from Eq. (19) that a numerical calculation of $\bar{\Phi}$ for Mars requires an estimate of $\tau_f$, so with that goal we write:

$$\tau_f = \frac{3.55 \cdot 10^3}{\beta} my, \qquad (56)$$

where $\beta$ is a number close to 1. For example, the range of values $3000 my < \tau_f < 4000 my$ is covered by $\sim 0.9 < \beta < \sim 1.2$. Hence, using Eq. (56) in Eq. (19), we obtain:

$$\bar{\Phi} = \frac{\beta 10^6}{D^{4.3}} (my)^{-1}, \qquad (57)$$

and thus the cumulative rate is:

$$\bar{\Phi}_C(\text{Mars}) = \frac{\beta 10^6}{3.3 D^{3.3}} (my)^{-1}. \qquad (58)$$

For instance, for $D = 20 km$ we obtain:

$$\Phi_C(\text{Mars}, 20km) \cong \frac{15\beta}{my} \approx \frac{15}{my}, \qquad (59)$$

which implies that the cumulative flux per area is, with $R_m$ being the Martian radius,

$$\frac{15/(4\pi R_m^2)}{my} \cong \frac{100 \times 10^{-9}}{(my) km^2}. \qquad (60)$$

The above results are considerably higher than the values for Earth from Eq. (47) and the results of Grieve and Shoemaker (1994) that are respectively :

$$2.8 \left[ \frac{1 \mp 0.50}{my} \right], \qquad (61)$$

$$\frac{(5.5 \mp 2.7) 10^{-9}}{(my) km^2}. \qquad (62)$$

Furthermore, for $D = 1km$, or, equivalently, impactor energies of around 1 to 2 megatons, we have, using Eq. (58),

$$\bar{\Phi}_C(\text{Mars}, 1km) = \frac{\beta}{3.3 y} \approx \frac{1}{3.3 y}. \qquad (63)$$

It is not surprising that the impact rate on Mars is larger than that on Earth, because of Mars' proximity to the asteroid belt; however, this shows that future Mars astronauts may have to deal with frequent damaging meteorite collisions. In particular, from the results above, we expect that Mars visitors spending a few years there will have a high probability of witnessing a megaton-type meteorite impact. Moreover, these impacts are likely to cause more damage on the surface than on our planet, due to the much lower atmospheric Martian density.

**7. Application of the Model to Earth's Cumulative Crater Number-Size Distribution**

Our model predicts, in accordance with Eqs. (13) and (12), that the number of craters with diameters between $D_i$ and $D_f$ and ages between $\tau_i$ and $\tau_f$ is given by the expression:

$$\widetilde{N} \equiv \int_{D_i}^{D_f} dD \int_{\tau_i}^{\tau_f} \Phi\{Exp[-\bar{C}\tau]\}d\tau. \tag{64}$$

Furthermore, assuming that:

$$\Phi = \frac{A_r}{D^m}, \tag{65}$$

$$\bar{C} = \frac{B}{D^p}, \tag{66}$$

we obtain from Eq. (64), as shown in Appendix D,

$$\widetilde{N} = \frac{A_r}{pB}(B\tau_i)^{-n}\{\Gamma[n, \frac{B\tau_i}{D_f^p}, \frac{B\tau_i}{D_i^p}] - (\frac{\tau_i}{\tau_f})^n \Gamma[n, \frac{B\tau_f}{D_f^p}, \frac{B\tau_f}{D_i^p}]\}, \tag{67}$$

where:

$$\Gamma[n,x,y] \equiv \int_x^y x^{n-1} e^{-x} dx, \tag{68}$$

is the generalized incomplete Gamma function, and:

$$n \equiv \frac{m-p-1}{p}. \tag{69}$$

The above equation for $\widetilde{N}$ contains the parameters $m, p, B$, and the amplitude $A_r$ which will be defined more precisely below. The values to be used for $m$ and $p$ are those determined for Mars, since this is strongly suggested by our previous arguments, observations and interpretation of the model. Also, we will write:

$$B = \frac{12 b1}{my}, \tag{70}$$

where $b1$ is a parameter that allows for the uncertainties in the value of $B$, which was determined, by the graphs in Figures (3) and (4), to be approximately:

$$B(g = 0.5) \cong \frac{11}{my}, \tag{71}$$

$$B(g = 0.1) \cong \frac{12}{my}, \tag{72}$$

where $g \equiv \tau_L/\tau_R$. Furthermore, $A_r$ is defined by:

$$\Phi_r = \frac{A_r}{D^m}, \tag{73}$$

where $\Phi_r$ is the flux of impacts on the Earth's surface chosen for the application of Eq. (67). In order to reduce the uncertainties due to undercounting in the crater data we will select the following regions for this study:

(a) Continental United States
(b) Canada up to the Arctic Circle latitude
(c) Europe
(d) Australia

The above area is considered to have well-counted crater data, particularly for $D > 20km$. $A_r$ is then the reduced flux amplitude corresponding to this region, which is proportional to the fraction of our planet covered by the above area, and thus we have:

$$A_r = A \frac{Area\ Under\ Consideration}{Earth's\ Surface\ Area}, \tag{74}$$

where $A$ is given, from Eq. (48), by:

$$A = (1.82)10^5[1 \mp 0.5]/my. \tag{75}$$

The crater data is taken from The Planetary and Space Science Centre. The area under consideration is approximately 30 million $km^2$, with an uncertainty well below that of $A$ so that we can write:

$$A_r \cong A(5.9 10^{-2}) = (1.07)10^4 a1, \tag{76}$$

where $a1$ represents the uncertainties in $A_r$, and is approximately given by:

$$a1 \cong [1 \mp 0.5]/my. \tag{77}$$

Therefore we can write the theoretical $\widetilde{N}$ with no free parameters, except for the uncertainties in $A_r$ and $B$ that are reflected in $a1$ and $b1$. We do this first in Table III and Figure (5) for craters with $D \geq 20km$ and cumulative age starting with $\tau = 1my$ up to $\tau = 2,000my$. Furthermore, we put $\tau_f = 2,500my$ and $D_f = 300km$, since all craters in the field of study are within this bin size. This theoretical curve, $\widetilde{N}(\tau)$, is compared with the corresponding observational data, allowing a Mathematica program to do a fitting only on the values of $a1$ and $b1$. The values of $a1$ and $b1$ arising from this best fitting are $a1 = 0.80/my$ and $b1 = 1.68$, which are within the expected uncertainties, and the very good agreement between theory and observations is noteworthy.

On the other hand, we also compared theory and observation in Table IV and Figure (6), where now $\widetilde{N}$ accumulative represents the number of craters of all ages, $1my \leq \tau \leq 2,500$, with diameters greater than or equal to $D$. Again, the theoretical $\widetilde{N}(D)$ is in very good agreement with the observations for $D \geq \sim 20km$, although not so good for $D \leq \sim 20km$, which is as expected due to the undercounting, which has already been mentioned, of craters of these sizes. Note that the slope of these accumulative curves is a function of $D$, and therefore it is not simply characterized by exponents as was the case in Figure (2) for Mars.

Perhaps it should be remarked that, in order to understand the similarities between theory and observation for the Earth's crater data when considering very low numbers of craters, a probabilistic approach to our model predictions is necessary. This statistical view could be similar to how the results of opinion polls are justified when based on only a small fraction of the total population: if the small sample of the poll is a well-chosen representative of the whole, then the errors remain low.

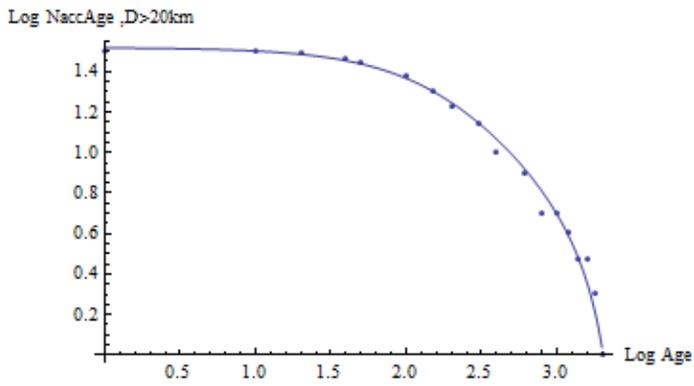

FIGURE (5): $Log[\widetilde{N}] vs Log[\tau \equiv Age]$, for all Diameters $D \geq 20km$. See Table III.

**Table III**

| τ(my) | $\widetilde{N}[\tau, D \geq 20km\ ]$ | Observation |
|---|---|---|
| 1 | 33.14 | 33 |
| 10 | 32.00 | 32 |
| 20 | 30.80 | 31 |
| 40 | 28.62 | 29 |
| 50 | 27.62 | 28 |
| 100 | 23.40 | 24 |
| 150 | 20.24 | 20 |
| 200 | 17.80 | 17 |
| 300 | 14.20 | 13 |
| 400 | 11.70 | 10 |
| 600 | 8.50 | 8 |
| 800 | 6.50 | 5 |
| 1000 | 5.00 | 5 |
| 1200 | 3.89 | 4 |
| 1400 | 2.99 | 3 |
| 1600 | 2.25 | 3 |
| 1800 | 1.62 | 2 |
| 2000 | 1.08 | 1 |

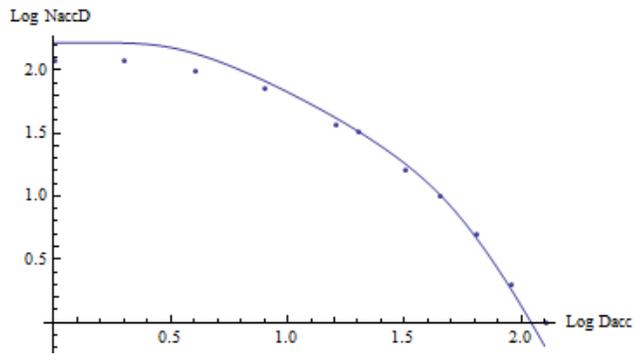

FIGURE (6): $[Log[\widetilde{N}]$ vs $Log[D \equiv D_{Acc}]$, for all ages between $1my \leq \tau \leq 2,500my$. See Table IV below.

**Table IV**

| D | $\widetilde{N}[D, 1my \leq \tau \leq 2,500my\ ]$ | Observation |
|---|---|---|
| 1 | 166.00 | 121 |
| 2 | 165.00 | 118 |
| 4 | 137.00 | 99 |
| 8 | 82.60 | 72 |
| 16 | 42.40 | 37 |
| 20 | 33.14 | 33 |
| 32 | 18.18 | 16 |
| 45 | 10.37 | 10 |
| 64 | 4.79 | 5 |
| 91 | 1.82 | 2 |
| 128 | 0.62 | 1 |

# Appendix A

Consider the following initial way of writing $N(D)$ (see Appendix B):

$$N = (\overline{\Phi/C})\{1 - Exp[-\bar{C}\tau]\},  \qquad \text{A.1}$$

where

$$(\overline{\Phi/C}) \equiv \frac{\int_0^\tau (\Phi(\tau`)/C)\{Exp[-\int_0^{\tau`} C(\tau``)d\tau``]\}Cd\tau`}{\int_0^\tau \{Exp[-\int_0^{\tau`} C(\tau``)d\tau``]\}Cd\tau`}, \qquad \text{A.2}$$

$$\bar{C} \equiv \frac{\int_0^\tau C d\tau`}{\tau}. \qquad \text{A.3}$$

We note that $\bar{C}$ is the time average of $C$ and $(\overline{\Phi/C})$ is the weighted average of $\Phi/C$, with weight:

$$\int_0^\tau \{Exp[-\int_0^{\tau`} C(\tau``)d\tau``]\}Cd\tau`. \qquad \text{A.4}$$

From physical considerations we expect that, in the limits $D \to \infty$ ($D \to 0$), we will have $\bar{C} \to 0$ ($\bar{C} \to \infty$), and therefore, from Eq. (A.1), with $\tau = \tau_f$, we see that:

$$N \to (\overline{\Phi/C})\bar{C}\tau_f, \quad D \to \infty, \bar{C} \to 0, \qquad \text{A.5}$$

$$N \to (\overline{\Phi/C}), \quad D \to 0, \bar{C} \to \infty. \qquad \text{A.6}$$

On the other hand, we also know that in the limit $D \to \infty, \bar{C} \to 0$ craters are conserved, and consequently

$$N = \int_0^{\tau_f} \Phi d\tau \equiv \bar{\Phi}\tau_f, \quad D \to \infty, \bar{C} \to 0. \qquad \text{A.7}$$

Hence, comparing Eq. (A.5) with Eq. (A.7), we obtain:

$$(\overline{\Phi/C}) = \frac{\bar{\Phi}}{\bar{C}} \qquad D \to \infty, \bar{C} \to 0. \qquad \text{A.8}$$

Moreover, in the limit $D \to 0, \bar{C} \to \infty$, we expect that the rate of production, $\Phi$, is equal to the rate of elimination, $NC$, hence, keeping $N$ constant (i.e. saturation), and we thus have

$$\Phi = NC; \quad D \to 0, \bar{C} \to \infty, \qquad \text{A.9}$$

or, since $N$ is now constant,

$$N = \frac{\bar{\Phi}}{\bar{C}}; \quad D \to 0, \bar{C} \to \infty. \qquad \text{A.10}$$

Comparing Eq. (A.10) with Eq. (A.6) we again obtain:

$$(\overline{\Phi/C}) = \frac{\bar{\Phi}}{\bar{C}}, \quad D \to 0, \bar{C} \to \infty. \qquad \text{A.11}$$

Therefore we have shown in Eqs. (A.8) and (A.11) that:

$$(\overline{\Phi/C}) = \frac{\bar{\Phi}}{\bar{C}}, \quad D \to 0 \; or \; D \to \infty. \qquad \text{A.12}$$

Although the identification in Eq. (A.12) is only in the limits $D \to 0 \; or \; D \to \infty$, this however corresponds to the part of the crater data that is described respectively by the two straight lines in Eqs. (27) and (25) and in the log-log plot in Figure (2), which essentially comprise all data points.

## Appendix B

We are going to show that the equation:

$$N = \int_0^\tau \Phi(\tau`)\{Exp[-\int_0^{\tau`} C(\tau``)d\tau``]\}d\tau`, \qquad \text{B.1}$$

can be written as Eq. (A.1):

$$N = (\overline{\Phi/C})\{1 - Exp[-\bar{C}\tau]\}. \qquad \text{B.2}$$

where

$$(\overline{\Phi/C}) \equiv \frac{\int_0^\tau (\Phi(\tau`)/C)\{Exp[-\int_0^{\tau`} C(\tau``)d\tau``]\}Cd\tau`}{\int_0^\tau \{Exp[-\int_0^{\tau`} C(\tau``)d\tau``]\}Cd\tau`}. \qquad \text{B.3}$$

$$\bar{C} \equiv \frac{\int_0^\tau Cd\tau`}{\tau}. \qquad \text{B.4}$$

Note that $\bar{C}$ is the time average of $C$ and $(\overline{\Phi/C})$ is the weighted average of $\Phi/C$, with weight:

$$\int_0^\tau \{Exp[-\int_0^{\tau`} C(\tau``)d\tau``]\}Cd\tau`. \qquad \text{B.5}$$

First, by equating Eq. (B.1) to Eq. (B.2) and using Eq. (B.3), we see that the demonstration is equivalent to showing that:

$$\int_0^\tau \{Exp[-\int_0^{\tau`} C(\tau``)d\tau``]\}Cd\tau` = \{1 - Exp[-\bar{C}\tau]\}. \qquad \text{B.6}$$

To this end, consider the elementary integral:

$$\int_0^y e^{-y`} dy` = 1 - e^{-y}, \qquad \text{B.7}$$

so that, if we define

$$y` \equiv \int_0^{\tau`} Cd\tau``, \qquad \text{B.8}$$

$$dy` = C(\tau`)d\tau`, \qquad \text{B.9}$$

we obtain, substituting Eqs. (B.8) and (B.9) in Eq. (B.7),

$$\int_0^\tau \{Exp[-\int_0^{\tau`} C(\tau``)d\tau``]\}Cd\tau` = 1 - Exp[-\int_0^\tau Cd\tau`], \qquad \text{B.10}$$

which, after using the definition in Eq. (B.4), is Eq. (B.6).   QED.

## Appendix C: Derivation of Equation (36)

The average of the diameters of craters of age $\tau \mp \Delta\tau$ observed today is given by:

$$\overline{D} = \frac{\int_0^\infty D N(D,\tau_L,\tau_R) dD}{\int_0^\infty N(D,\tau_L,\tau_R) dD} \quad , \qquad \text{C.1}$$

where

$$N(D,\tau_L,\tau_R) = \int_{\tau_L}^{\tau_R} \Phi(D,\tau) Exp[-\bar{C}\tau] d\tau \quad . \qquad \text{C.2}$$

If $\Phi$ and $\bar{C}$ are independent of $\tau$, we have, after integrating Eq. (C.2):

$$N(D,\tau_L,\tau_R) = \frac{\Phi}{\bar{C}}(e^{-\bar{C}\tau_L} - e^{-\bar{C}\tau_R}), \qquad \text{C.3}$$

and furthermore if:

$$\Phi = \frac{A}{D^m}, \qquad \text{C.4}$$

$$\bar{C} = \frac{B}{D^p}, \qquad \text{C.5}$$

we obtain from Eq. (C.3):

$$N(D,\tau_L,\tau_R) = \frac{A}{B} D^{-m`}[e^{-x} - e^{-\bar{x}}], \qquad \text{C.6}$$

where, with $g = \tau_L/\tau_R$,

$$m` = m - p, \qquad \text{C.7}$$

$$x \equiv \frac{B\tau_L}{D^p}, \qquad \bar{x} \equiv \frac{\bar{B}\tau_L}{D^p}, \quad \bar{B} \equiv \frac{B}{g}. \qquad \text{C.8}$$

Therefore, substituting Eq. (C.6) in Eq. (C.1) we obtain:

$$\overline{D} = \frac{\int_0^\infty D^{-m`+1}[e^{-x}-e^{-\bar{x}}]dD}{\int_0^\infty D^{-m`}[e^{-x}-e^{-\bar{x}}]dD}. \qquad \text{C.9}$$

Now, we can write, for any constant $k$, (see Appendix D for details):

$$\int_0^\infty D^{-k} e^{-x} dD = \frac{(B\tau_L)^{\frac{1-k}{p}}}{p} \int_0^\infty x^{\frac{k-1-p}{p}} e^{-x} dx, \qquad \text{C.10}$$

$$\int_0^\infty D^{-k} e^{-\bar{x}} dD = \frac{(\bar{B}\tau_L)^{\frac{1-k}{p}}}{p} \int_0^\infty \bar{x}^{\frac{k-1-p}{p}} e^{-\bar{x}} d\bar{x}. \qquad \text{C.11}$$

However, we have that $\int_0^\infty x^{\frac{k-1-p}{p}} e^{-x} dx = \Gamma\left(\frac{k-1}{p}\right) =$ the complete Gamma Function, and thus we obtain from Eq. (C.9), after using Eqs. (C.10) and (C.11),

$$\overline{D} = \frac{((B\tau_L)^{\frac{2-m`}{p}} - (B\tau_{L/g})^{\frac{2-m`}{p}}) \Gamma\left(\frac{m`-2}{p}\right)}{((B\tau_L)^{\frac{1-m`}{p}} - (B\tau_{L/g})^{\frac{1-m`}{p}}) \Gamma\left(\frac{m`-1}{p}\right)} = (B\tau_L)^{\frac{1}{p}} \alpha 1, \qquad \text{C.12}$$

where

$$\alpha 1 = \frac{\Gamma((m`-2)/p)}{\Gamma((m`-1)/p)} \frac{[1-g^{(m`-2)/p}]}{[1-g^{(m`-1)/p}]}, \qquad \text{C.13}$$

thus arriving at Eq. (36).

## Appendix D: Deriving Equation (67)

Consider Eq. (64)

$$\tilde{N} = \int_{D_i}^{D_f} (\mathrm{d}D)\, N(D, \tau_i, \tau_f) \equiv \int_{D_i}^{D_f} dD \int_{\tau_i}^{\tau_f} \Phi\{Exp[-\bar{C}\tau]\} d\tau, \quad \text{D.1}$$

and assume that:

$$\Phi = \frac{A_r}{D^m}, \quad \text{D.2}$$

$$\bar{C} = \frac{B}{D^p}, \quad \text{D.3}$$

We then obtain, after integrating first with respect to $\tau$,

$$\tilde{N} = \int_{D_i}^{D_f} \left(\frac{A_r}{B} D^{-m`} [e^{-x} - e^{-\bar{x}}]\right) dD, \quad \text{D.4}$$

where:

$$m` = m - p, \quad \text{D.5}$$

$$x \equiv \frac{B\tau_i}{D^p}, \quad \bar{x} \equiv \frac{\bar{B}\tau_i}{D^p}, \quad \bar{B} \equiv \frac{B}{g}. \quad \text{D.6}$$

We now change the variable of integration in Eq. (D.4) from $D$ to $x$:

$$D = \left(\frac{B\tau_i}{x}\right)^{1/p}, \quad \text{D.7}$$

$$dD = -\left(\frac{1}{p}\right)(B\tau_i)^{1/p} x^{-1-1/p}) dx, \quad \text{D.8}$$

from which we obtain:

$$\int_{D_i}^{D_f} D^{-m`} Exp[-x] dD = \left(\frac{1}{p}\right)(B\tau_i)^{(1-m`)/p} \int_{x_f}^{x_i} x^{\frac{(m`-1-p)}{p}} Exp[-x] dx = \left(\frac{1}{p}\right)(B\tau_i)^{-n} \Gamma[n, x_f, x_i], \quad \text{D.9}$$

Where

$$n = (m` - 1)/p, \quad \text{D.10}$$

and we use the generalized Gamma function:

$$\Gamma[n, x_i, x_f] = -\Gamma[n, x_f, x_i] \equiv \int_{x_i}^{x_f} x^{n-1} Exp[-x] dx. \quad \text{D.11}$$

Likewise, we also have:

$$\int_{D_i}^{D_f} D^{-m`} Exp[-\bar{x}] dD = \left(\frac{1}{p}\right)(\bar{B}\tau_i)^{-n} \Gamma[n, \bar{x}_f, \bar{x}_i]. \quad \text{D.12}$$

Note that if $D_i = 0$ and $D_f = \infty$ then $x_i = \infty$ and $x_f = 0$, and if we identify $m`$ with $k$ in Eqs. (C.10) and (C.11), then Eqs. (D.9) and (D.12) become Eqs. (C.10) and (C.11). Substituting Eqs. (D.9) and (D.12) in Eq. (D.4), we find that:

$$\widetilde{N} = \frac{A_r}{Bp}\{(B\tau_i)^{-n}\Gamma[n, x_f, x_i] - (\bar{B}\tau_i)^{-n}\Gamma[n, \bar{x}_f, \bar{x}_i]\} \quad . \tag{D.13}$$

Thereby, with Eq. (D.6) and some algebra, we find the result in Eq. (67):

$$\widetilde{N} = \frac{A_r}{pB}(B\tau_i)^{-n}\{\Gamma[n, \frac{B\tau_i}{D_f^p}, \frac{B\tau_i}{D_i^p}] - (\frac{\tau_i}{\tau_f})^n \Gamma[n, \frac{B\tau_f}{D_f^p}, \frac{B\tau_f}{D_i^p}]\}. \tag{D.14}$$